\documentstyle[prd,twocolumn,aps,epsf]{revtex}
\newcommand{\singlefig}[2]{
\begin{center}
\begin{minipage}{#1}
\epsfxsize=#1
\epsffile{#2}
\end{minipage}
\end{center}}
\newcommand{\segmentfig}[3]{
\begin{minipage}{#1}
\epsfxsize=#1
\epsffile{#2}
\begin{center}
{\small \mbox{#3}}
\end{center}
\end{minipage}}
\newcommand{\gsim}{\mbox{\raisebox{-1.ex}{$\stackrel
     {\textstyle>}{\textstyle\sim}$}}}

\begin{document}
\draft
\title{Dyonic BIon black hole in string inspired model}
\author{Takashi Tamaki
\thanks{electronic
mail:tamaki@gravity.phys.waseda.ac.jp}}
\address{Department of Physics, Waseda University,
Ohkubo, Shinjuku, Tokyo 169-8555, Japan}
\author{Takashi Torii
\thanks{electronic mail:torii@resceu.s.u-tokyo.ac.jp}}
\address{Research Center for the Early Universe, 
University of Tokyo, Hongo, Bunkyo, Tokyo 113-0033, Japan
\\ and \\
Advanced Research Institute for Science and Engineering,
Waseda University, Ohkubo, 
Shinjuku, Tokyo 169-8555, Japan}
\date{\today}
\maketitle
\begin{abstract}
We construct static and spherically symmetric particle-like 
and black hole solutions with magnetic and/or electric charge 
in the Einstein-Born-Infeld-dilaton-axion system, which is a 
generalization of the Einstein-Maxwell-dilaton-axion (EMDA) system 
and of the Einstein-Born-Infeld (EBI) system. They have remarkable 
properties which are not seen for the corresponding solutions in the EMDA  
and the EBI system. If solutions do not have both magnetic and electric charge, 
the axion field becomes trivial. 
In the electrically charged case, 
neither the extreme nor the BPS saturated solutions exist. Although we can 
take the zero horizon radius limit for any Born-Infeld (BI) parameter $b$, 
there is no particle-like solution. 
In the magnetically 
charged case, the extreme solution {\it does} exist for the critical BI 
parameter (or charge) $\sqrt{b}Q_m=1/2$. The critical BI parameter divides the solutions 
qualitatively. For $\sqrt{b}Q_m<1/2$, there exists a particle-like solution
for which the dilaton field is finite everywhere, while the no particle-like solution 
exists and the solution in the $r_h \to 0$ limit becomes naked for $\sqrt{b}Q_m>1/2$. 
Though there is an extreme solution, the BPS saturated solution 
does not exist in this case. 
When the solutions have both magnetic and electric charge, 
we obtain the nontrivial axion field which plays an important role particularly for 
small black holes. Thermodynamical properties and the
configuration of the dilaton field approach
those in the magnetically charged case in the 
zero horizon limit, although gravitational mass does not. This is related to 
the nontrivial behavior of the axion field.  
We can prove that there is no inner horizon and that the global structure is 
the same as the Schwarzschild black hole in any charged case.
\end{abstract}

%
\section{Introduction}
%
The pioneering theory of the non-linear electromagnetic field was
formulated by Born and Infeld (BI) in 1934\cite{Born}. 
Surprisingly, it has been shown that BI action arises in string-generated
corrections if one considers the coupling of an 
Abelian gauge field to an open bosonic string or 
an open superstring\cite{Frad}. Moreover, the world volume 
action of a D-brane is described by a kind of non-linear 
BI action in the weak string coupling limit\cite{brane}. 
In this respect, there have been many investigations about BI action. 
For example, a particle-like solution (BIon) was constructed and
its relation to the fundamental strings attached to the brane 
was discussed\cite{BIon}. 
The BI action in a constant background of the
Kalb-Ramond potential $B_{\mu\nu}$ turns out to be described by gauge theory 
in a flat noncommutative spacetime\cite{Witten}. BIon in such a spacetime 
was also discussed recently\cite{Mateos}. 
Modification of light propagation is also important and the general geometric 
aspects including the BI action was discussed in Ref.~\cite{Novello}. 
Fluctuation around a nontrivial solution of BI action have a limiting speed 
given not by the Einstein metric but by the Boillat metric\cite{Boillat} which is 
conformal to the open string metric\cite{Gary}. In general, characteristics of 
the wave exhibits bi-refrigence. 
The BI action is an exceptional theory which is free from this. 

Some extensions of BI action to the non-Abelian gauge field has been 
considered, although it is not determined uniquely because of ambiguity in taking 
the trace of internal space\cite{Tsey}. Classical glueball solutions 
which were prohibited in the standard Yang-Mills theory were reported in Ref. \cite{Gal}. 
They were also considered in the non-Abelian BI action with symmetrized trace and 
the dependence of their properties on the method of taking the trace 
and on gravity were discussed in Ref. \cite{Dya}. 
Earlier work to consider Einstein gravity in the Abelian BI action was first 
done by Demianski\cite{Demi}. Then self-gravitating particle-like solutions (EBIon)\cite{foot2} 
and their black hole solutions (EBIon black hole) were calculated analytically under a 
static spherically symmetric ansatz\cite{Demi,Oliveira}. The extension to consider 
higher curvature gravity was considered in Ref.~\cite{Wil}. 

The non-linearity of the electromagnetic field may bring remarkable properties 
such as nonsingular black hole solutions satisfying the weak energy 
condition reported in \cite{Gar,Eloy}. They are distinct from Bardeen black holes 
in the point that they appear as solutions in the Einstein equation with a nonvanishing 
matter field\cite{Bardeen}. Although the singularity appears in the BI model, we can 
find an intrinsic difference from the Reissner-Nortstr\"om (RN) black hole concerning 
causal structure and black hole thermodynamics. 

Since the BI system, however, has a string theoretical origin, 
it should include the dilaton 
and the antisymmetric Kalb-Ramond tensor field, which we call the axion field, as
formulated  in Ref.~\cite{Callan}. It is a direct extension of the 
Einstein-Maxwell-dilaton-axion (EMDA) system, where the famous black hole solution 
was found with the vanishing axion (i.e., in the EMD system) 
by Gibbons and Maeda, and independently 
by Garfinkle, Horowitz and Strominger (GM-GHS)\cite{GM-GHS}. 
The black hole solutions including the axion field were also reported in 
Ref.~\cite{axion,Wilczek}. In this paper, 
we investigate the particle-like and black hole solutions 
in the Einstein-Born-Infeld-dilaton-axion (EBIDA) system, which we call dilatonic 
EBIon (DEBIon) and dilatonic EBIon black hole (DEBIon black hole), respectively.

We studied such solutions only when they have either 
magnetic or electric charge and clarified 
the effect of the dilaton field as a first step in the previous work\cite{TTpast}. 
The magnetically charged solution 
and electrically charged solution have different properties 
because of the absence of 
electric-magnetic duality\cite{Gibbons}. This system was modified to satisfy 
such duality\cite{Rasheed}, and solitons and black holes are also considered 
in that system\cite{Clement}.  
Since the magnetic and electric charges are, however, not taken into account
simultaneously in these analyses, 
the axion field becomes trivial by assuming a spherically 
symmetric ansatz. It is intriguing that the dyon solution with a 
trivial axion field has a different global structure and thermodynamical properties 
from the monopole case in the EMD system, while that with the 
nontrivial axion field has the same 
properties as the monopole case because of the SL($2$,R) duality 
in the EMDA system. Both BI systems are formulated, including the axion 
field, and we need to clarify their role. 
Motivated by these factors, we  consider 
the dyonic black hole with a nontrivial axion field. 
We use the original action considered in Ref.~\cite{Callan}. 
Though this action loses electric-magnetic duality, this 
may be restored if we consider other elements such as 
higher derivative terms of the field strength $F$. 

This paper is organized as follows. In Sec. II, we describe our model, 
basic equations and boundary conditions. We explain electrically charged solutions 
and magnetically charged solutions in Sec. III and IV, respectively. 
We observe that the thermodynamical properties are distinct in these two cases, 
which is caused by the effect of the dilaton field. In Sec.~V, 
we consider a dyon solution 
with a nontrivial axion field. This has various interesting properties which were not 
seen in the monopole cases. In particular, even when the ratio of the electric charge to
the magnetic charge is large,
the solution does not approach the electrically charged one 
because of the nontrivial axion configuration. 
Moreover, when we consider the evaporation process of the black hole, it 
approaches the extreme solution for some mass scale and 
has a very low Hawking temperature.  
When the mass of the black hole shifts from this mass scale, 
it exhibits abrupt growth in temperature which eventually diverges 
in the zero horizon limit. We denote concluding remarks and future work in Sec. VI. 
Throughout this paper, we use the units $c=G=\hbar =1$.

\section{Model and Basic Equations}

We start with the following action\cite{notes};
\begin{eqnarray}
& & S=\int d^{4}x\sqrt{-g}\left[
\frac{R}{2\kappa^{2}}-\frac{  (\nabla\phi )^{2} }{\kappa^{2}}
-\frac{1}{24\kappa^{2}}e^{-4\gamma\phi}H^{2}
+L_{BI}\right],
\label{EBID}
\nonumber
\\ & & 
\end{eqnarray}
where $\kappa^{2} := 8\pi $ and  $\gamma$ is the
coupling constant of the dilaton field $\phi$.
The three rank antisymmetric tenser field is 
expressed as, $H=dB+\frac{1}{4}A\wedge F$.
 
$L_{BI}$ is the BI part of the Lagrangian which is written as
\begin{eqnarray}
L_{BI}=\frac{b e^{2\gamma\phi}}{4\pi}
\left\{
1-\sqrt{1+\frac{e^{-4\gamma\phi}}{2b}P
-\frac{e^{-8\gamma\phi}}{16b^{2}}Q^{2}    }
\right\}  \ ,
\label{matter}
\end{eqnarray}
where $P:= F_{\mu\nu}F^{\mu\nu}$ and 
$Q:= F_{\mu\nu}\tilde{F}^{\mu\nu}$.
A tilde denotes the Hodge dual. 
We can rewrite the three rank antisymmetric tenser field 
using a single pseudo scalar field $a$ (the axion field) as 
\begin{eqnarray}
H^{2}
=6e^{8\gamma\phi}(\nabla a)^{2}-12e^{4\gamma\phi}aQ . 
\label{Hfield}
\end{eqnarray}
When we examine the electric and the magnetic monopole
cases separately, the term $Q$ vanishes and eventually 
the axion field becomes trivial by assuming a spherically 
symmetric ansatz. But for the dyonic case, which has both electric and 
magnetic charges, the axion becomes nontrivial. We will treat these cases 
systematically. 
The BI parameter $b$ has the physical interpretation of a 
critical field strength. In the string theoretical context,
$b$ is related to the inverse string tension 
$\alpha'$ by $b^{-1}=(2\pi \alpha ')^{2}$.
Notice that the action (\ref{EBID}) is reduced to the EMDA system in the 
limit $b \rightarrow \infty$ and
to the EBI system with the massless field for $\gamma =0$ and $a=0$. 
Here, we concentrate on the case $\gamma =1$
which is predicted from superstring theory
and $\gamma=0$ for comparison.

We consider the metric of static and spherically symmetric,
\begin{eqnarray}
ds^{2}=-f(r) e^{-2\delta (r)}dt^{2}  
+f(r)^{-1}dr^{2}+r^{2}{d\Omega_2}^{2} \ ,
\label{metric}
\end{eqnarray}
where $f(r)=1-2m(r)/r$. 
The gauge potential has the following form: 
\begin{eqnarray}
A&=&-\frac{w_{1}(r)}{r}dt-w_{2}(r)\cos \theta d\varphi.
\label{elemag}
\end{eqnarray}
From the BI equation, we obtain that $w_{2}\equiv Q_{m}$ 
is constant and
\begin{eqnarray}
\left(\frac{w_{1}}{r}\right)'
=-YQ_{e}e^{2\phi-\delta}
\left[r^4+\frac{(YQ_{e})^{2}}{b}\right]^{-\frac12},
\label{ele2}
\end{eqnarray}
where $Q_{e}:= w_{1}(\infty)$, $q:=Q_{e}/Q_{m}$ and $Y:= 1-a/q$.
A prime denotes a derivative with respect to $r$. 
From this equation, the electric field $E_r=-(w_1/r)'$
does not diverge but takes a finite value at the origin.
The maximum value is $E_r=\sqrt{b}$
in the EBI system. As we will see, the electric field
vanishes at the origin in the EBIDA system
by the nontrivial behavior of $\phi$.
The potential $w_1$ is formally expressed as 
\begin{eqnarray}
w_{1}
=-r\int^r_0 YQ_{e}e^{2\phi-\delta}
\left[r^4+\frac{(YQ_{e})^{2}}{b}\right]^{-\frac12}dr,
\end{eqnarray}
where we put $rw_1(0)=0$ without loss of generality.
By the above ans\"{a}tze the basic equations 
with $\gamma=1$ are written as follows.
\begin{equation}
m'=-U+\frac{r^2}{2}f 
\left[(\phi')^{2}+\frac{e^{4\phi}}{4}(a')^{2}\right] ,
\label{m1} 
\end{equation}
\begin{equation}
\delta'=-r\left[
(\phi')^{2}+\frac{e^{4\phi}}{4}(a')^{2}\right] ,
\label{d1}   
\end{equation}
\begin{equation}
\phi ''=-\frac{2}{r}\phi'+\frac{e^{4\phi}}{2}(a')^{2}
-\frac{2}{f}
\left[\left(\frac{m}{r}+U\right)\frac{\phi'}{r}
-X \right] ,
\label{p1} 
\end{equation}
\begin{equation}
a''=-2a'\left(\frac{1}{r}+2\phi '\right)
-\frac{2}{f}
\left[\left(\frac{m}{r}+U\right)\frac{a'}{r}+2V\right] ,
\label{a1}
\end{equation}
where
\begin{equation}
U :=   br^{2}e^{2\phi}\left\{1-\sqrt{
\left(1+\frac{Q_{e}^{2}Y^{2}}{br^{4}}\right)
\left(1+\frac{e^{-4\phi}Q_m^2 }{br^{4}}\right)
}
\right\},
\label{BI}  
\end{equation}
\begin{equation}
V:=\frac{e^{-2\phi}Q_{m}Q_{e}Y}{  
r^{2}\sqrt{r^{4}+\frac{(YQ_{e})^{2}}{b}}
},
\label{KETTA} 
\end{equation}
\begin{equation}
X:=  be^{2\phi}
\left[\sqrt{\frac{br^4+(Q_{e}Y)^2}{br^4+Q_m^2e^{-4\phi}}}
\left(1+\frac{V^{2}}{b^{2}}\right)-1
\right].  
\label{DAHLIA}    
\end{equation}
Note that by introducing the dimensionless variables
$\bar{r}:= \sqrt{b}r$, $\bar{m}:= \sqrt{b}m$,
$\alpha_{e}:= \sqrt{b}Q_{e}$  and
$\alpha_{m}:= \sqrt{b}Q_{m}$,
we find that the parameters of the equation system are
$\alpha_{e}$ and $\alpha_{m}$.
As we already explained, since 
this system loses electric-magnetic duality, when we 
obtain the black hole solution with 
electric charge in the above action, we can not transform it to 
the solution with
magnetic charge by duality.

The boundary conditions at spatial infinity to satisfy the 
asymptotic flatness are 
\begin{eqnarray}
& m(\infty) =: M=const.,\ \ \delta (\infty)=0, &
\nonumber
\\ 
& \phi(\infty)=0,\ \ a(\infty)=0. &
\label{atinf}
\end{eqnarray}
We also assume the existence of a regular 
event horizon at $r=r_{h}$ for the DEBIon black hole. 
So we have
\begin{equation}
m_{h}=\frac{r_h}{2},\;\; \delta_h< \infty ,
\;\;  \phi_h<\infty ,\;\;  
\label{mth}  
\end{equation}
\begin{equation}
\phi'_{h}=\frac{2r_hX_h}{1+2U_h}, \;\;\;
a_{h}'=\frac{-4r_{h}V_{h}}{1+2U_{h} }.
\label{ath}
\end{equation}
The variables with subscript $h$ mean that they are evaluated 
at the horizon. We will obtain the black hole 
solution numerically by determining $a_{h}$ and $\phi_{h}$ 
iteratively to satisfy these conditions. 

To determine the boundary values of $\phi_{h}$ and $a_{h}$, 
it is convenient to rewrite the field equations (\ref{p1}) and 
(\ref{a1}) as follows, 
\begin{equation}
(fe^{-\delta}r^{2}\phi ')'=\frac{1}{2}fr^{2}e^{4\phi -\delta}(a')^{2}
+2r^{2}e^{-\delta}X ,
\label{p1p} 
\end{equation}
\begin{equation}
(fe^{4\phi-\delta}r^{2}a ')'=-4r^{2}e^{4\phi-\delta}V . 
\label{a1p}
\end{equation}
We first concentrate on the equation of the axion field. 
When the solution does not have both electric and magnetic charges, 
$A:=fe^{4\phi-\delta}r^{2}a '=const.$ by (\ref{a1p}). Since $A=0$ at 
the horizon, $a(r)'\equiv 0$ can be derived which means $a\equiv 0$ 
to satisfy $a(\infty )=0$. In the dyon case, 
since the difference of the sign of $Q_{m}Q_{e}$ only affects the sign of $a$ 
as is seen in Eqs. (\ref{m1})-(\ref{a1}), 
we restrict $Q_{m}Q_{e}>0$ below without loss of generality. 
Assuming $a_{h}>q$, we obtain $A'>0$ that follows $A>0$ outside the horizon. 
Thus, $a$ increases monotonically and cannot satisfy $a(\infty)=0$. 
Assuming $a_{h}<q$, we obtain $A'<0$ that follows $A<0$ outside the horizon. 
Thus, $a$ decreases monotonically and hence 
$0<a_{h}<q$ to satisfy $a(\infty)=0$. 
In this case, since we have $A>0$ 
(nonzero) inside the horizon, $f$ cannot become zero. 
So there is no inner horizon for the dyon case. 

We consider the equation of the dilaton field (\ref{p1p}). 
For the electrically charged case, 
we obtain $(fe^{-\delta}r^{2}\phi ')'>0$ 
because $X(r\neq 0)>0$ and $a(r)'\equiv 0$. 
Thus $C:=fe^{-\delta}r^{2}\phi '<0$ and $C>0$ inside and 
outside the horizon, respectively. 
So $\phi '>0$ everywhere and $f$ cannot become zero inside the horizon, 
i.e., there is no inner horizon. 
To satisfy $\phi (\infty )=0$, we should choose as $\phi_{h}<0$. 
We can discuss the magnetically charged case in the same way and 
find that $\phi_{h}>0$ and that there is no inner horizon. 
For the dyonic case, however, the sign of $X$ cannot be fixed  
since it depends not only on the charge ratio $q$ 
but also on the values $a_{h}$ and $\phi_{h}$ in a complex relationship.  

We obtained solutions numerically. 
For the $b\to\infty$ case, we have a GM-GHS solution. This does not 
guarantee an existence of a solution for finite $b$. 
First, we show the necessary condition for the existence of a 
solution. To satisfy $m'_{h}<1/2$, we have 
\begin{eqnarray}
&&4bQ_{e}^{2}Y_{h}^{2}e^{4\phi_{h}}-4br_{h}^{2}e^{2\phi_{h}}
+4bQ_{m}^{2}  \nonumber  \\
&&-1+\frac{4Q_{e}^{2}Q_{m}^{2}Y_{h}^{2}}{r_{h}^{4}}:=K<0 .
\label{condition1}
\end{eqnarray}
If we see the equation $K=0$ as the second order equation 
for $e^{2\phi_{h}}$, following 
condition will be required from its discriminant.
\begin{eqnarray}
br_{h}^{4}>Q_{e}^{2}Y_{h}^{2}\left[4Q_{e}^{2}
\left(b+\frac{Q_{e}^{2}Y_{h}^{2}}{r_{h}^{4}}\right)-1\right] . 
\label{condition2}
\end{eqnarray}
Thus, for the existence of the solution in the $r_{h}\to 0$ 
limit, $Y_{h}/r_{h}^{2}$ should approach finite value 
($0$ in our calculation) in this limit, which becomes important 
in Sec. V. Though, (\ref{condition1}) is a necessary condition, 
when the conditions (\ref{condition1}), 
(\ref{mth}), and (\ref{ath}) are satisfied, 
our calculation shows that the matter and the gravitational fields 
satisfy asymptotically flatness. 
This is plausible since the EBIDA system approaches the EMDA system 
for $r\to \infty$. 

\section{Electrically charged solution}
\begin{figure}
\begin{center}
\singlefig{10cm}{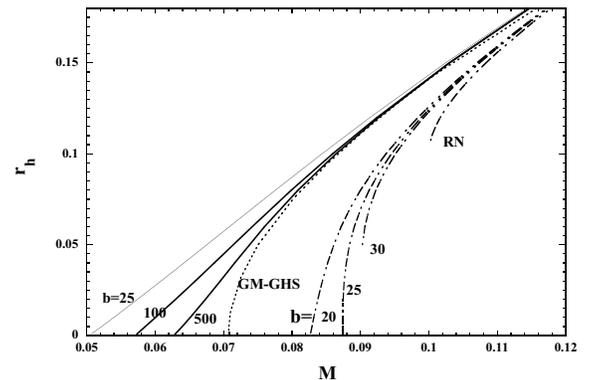}
\caption{$M$-$r_h$ diagram of electrically charged
DEBIon black holes (solid lines) with 
electric charge $Q_e =0.1$ and $b=25$, $100$ and $500$. 
GM-GHS and EBIon black holes are 
also plotted by a dotted line and dot-dashed lines, respectively. 
\label{M-rhele} }
\end{center}
\end{figure}

First, we investigate the electrically charged case.   
Before proceeding to the DEBIon black hole, we briefly review
the solutions in the EBI system ($\phi \equiv a \equiv 0$).
In the $b\to \infty$ limit, the EBI system reduces to 
the Einstein-Maxwell system. For finite $b$, we obtain EBIon and its black hole 
solutions analytically\cite{Demi,Oliveira}. 
We plot the $M$-$r_h$ relation in Fig.~\ref{M-rhele} by dot-dashed lines.
The solution branches are divided qualitatively by 
$\alpha_e=\alpha^{\ast}:=1/2$. For $\alpha_e >\alpha^{\ast}$, 
there is a special value $M_0$
of which the analytic form is seen in Ref.~\cite{Oliveira}.
For $M<M_0$ the black hole and inner horizon
exist as the RN black hole while
only the black hole horizon exists for $M \geq M_0$.
The minimum mass solution in each branch corresponds to the
extreme solution. On the other hand, all the black hole
solutions have only one horizon and the global structure
is the same as the Schwarzschild black hole
for $\alpha_e <\alpha^{\ast}$. In this case, there exists 
EBIon solution with no horizon in the $r_h \to 0$ limit. 
Note however that,  since $m'(0) = \alpha_e$,
it has the conical 
singularity at the origin, which is the characteristic
feature of the self-gravitating BIons.
For $\alpha_e =\alpha^{\ast}$, the extreme solution is 
realized in the $r_h \to 0$ limit. Although Demianski called it 
an electromagnetic geon \cite{Demi} which is regular everywhere,
it has a conical singularity like the other EBIons. 

Next we turn to the $\gamma=1$ case.
In the $b\to \infty$ limit, i.e., in the EMD system,
the GM-GHS solution exists. The GM-GHS solution has three global charges, 
mass $M$, the electric charge $Q_e$ and the dilaton charge $\Sigma$.
The dilaton charge depends on the former two, hence it is classified 
as a secondary charge\cite{Bizon}. For the GM-GHS solution, 
it is expressed as 
\begin{eqnarray}
\Sigma =-\frac{Q_{e}^{2}}{2M} .
\label{dchargeGM}
\end{eqnarray}
There is no particle-like solution in this system.

For the finite value of $b$, we can not find the analytic
solution and have to use numerical analysis. 
Since there is no non-trivial dilation configuration
in the $Q_e=0$ case, the dilaton hair is again the secondary
hair in this system.
We  plot the $M$-$r_h$ relation of 
the DEBIon black hole in Fig.~\ref{M-rhele} by solid lines.
We can find that all branches reach to $r_h=0$
in contrast to the EBIon case.
We examine the existence of the extremal solution. 
It has a degenerate horizon, where 
$2m'=1$ is realized. 
Hence, by Eq.~(\ref{m1})
\begin{eqnarray}
\alpha_{e}=\alpha_e^{ext}
:=e^{-2\phi_{h}}\sqrt{\frac{1}{4}+br_h^2 e^{2\phi_h}}.
\label{degenerate}
\end{eqnarray}
Since $\phi_{h}=0$ in the EBI system, the extreme solutions exist 
for $\alpha_{e}\geq \alpha^{\ast}$. In this case, (\ref{degenerate}) is 
satisfied when 
$r_{h}=\sqrt{(\alpha_{e}^{2}-1/4)/b}$. For the EBID system, 
since $\phi_h$ diverges to minus infinity in the 
$r_h\to 0$ limit as we will see below, $\alpha_e ^{ext} \to \infty$ 
and no extreme solution exists for finite $\alpha_e$. 

As for the particle-like solution, we have to analyze this 
carefully. We employ a new function $\psi :=e^{2\phi}$ 
and expand the field variables as
\begin{eqnarray}
\psi = \sum_{\alpha, \; \beta}\psi_ {(\alpha, \beta)} 
r^{\alpha} (\ln r)^{\beta},
\;\;\;\;
m = \sum_{\gamma, \; \delta}m_ {(\gamma, \delta)} 
r^{\gamma} (\ln r)^{\delta}.
\label{expand}
\end{eqnarray}
Substituting them into the field equations and
evaluating the lowest order equations, we find
\begin{equation}
\phi \sim -\frac{1}{2} \ln(-4\sqrt{bQ_e^2}\ln r),  
\label{asymDEBIonp}  
\end{equation}
\begin{equation}
m \sim -\frac{r}{4\ln r}. 
\label{asymDEBIonm}
\end{equation}
The diverging behavior of the dilaton field is important. 
Though it is admissible as a particle-like solution in the 
sense of Ref. \cite{foot2}, conformal transformation 
back to the original string frame becomes singular. 
In this respect, we use the word a particle-like solution below 
for the solution which satisfies both the criterion \cite{foot2} 
and the finiteness of the dilaton field. 

\begin{figure}
\begin{center}
\singlefig{10cm}{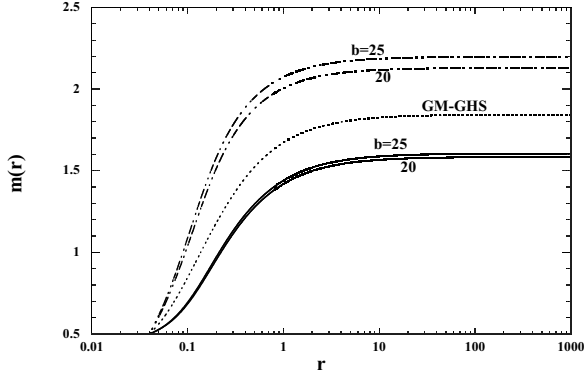}
\caption{The mass distribution of the electrically charged 
DEBIon black holes (solid lines) with $Q_e =0.1$, $r_{h}=0.04$ and $b=25$ and $20$.
GM-GHS and EBIon black holes cases are 
also plotted by a dotted line and dot-dashed lines, respectively. 
\label{r-mele}  }
\end{center}
\end{figure}

From Fig.~\ref{M-rhele}, we can find that the mass of the black hole becomes
small when we include the dilaton field into the system for any value
of BIon parameter $b$. The reason is clarified by examining the mass
distribution.
We show the mass distribution $r$-$m(r)$ of the electrically charged 
DEBIon black holes with $Q_e =0.1$, $r_{h}=0.04$ and $b=20$ and $25$
in Fig.~\ref{r-mele}. 
These configurations show how the dilaton field contributes to the mass function
when we compare it to the EBIon black hole case. 
Roughly speaking, the contribution comes from two factors: 
(i) the dilaton coupling prefactor $e^{2\phi}$ 
which appears in Eq.~(\ref{BI}) and (ii) the gradient term $(\phi')^{2}$. 
Since $\phi$ takes negative value in the electrically charged case, 
factor (i) reduces the gravitational constant effectively. 
On the other hand, the second factor makes positive contribution to the 
mass function. We can find that factor (i) overcomes factor (ii),
since gravitational mass is reduced compared with the one having no 
dilaton field. This effect  appears  particularly near the horizon
where the dilaton field largely deviates from zero.  

\begin{figure}
\begin{center}
\singlefig{10cm}{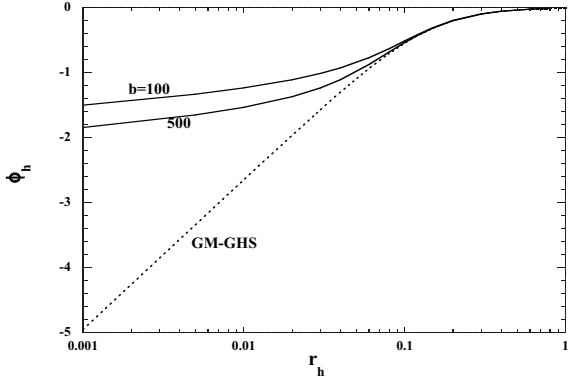}
\caption{$r_{h}$-$\phi_{h}$ relation of electrically charged 
DEBIon black holes (solid lines) with 
$Q_e =0.1$ and $b=100$ and $500$. 
The GM-GHS solution is plotted by a dotted line. 
This diagram shows the qualitative difference of the divergence rate
in the $r_{h}\to 0$ limit
between the DEBIon black hole and the GM-GHS solution. 
\label{rh-phihele}  }
\end{center}
\end{figure}

We show the relation between the horizon radius $r_{h}$ and 
$\phi_{h}$ in Fig. \ref{rh-phihele}. We can find qualitative 
difference of the scalar field in the $r_{h}\to 0$ limit between the DEBIon 
black hole and the GM-GHS solution. 
The dilaton field of  the GM-GHS solution diverges as
\begin{eqnarray}
\phi_{h} \sim \ln \left(\frac{Mr_{h}}{Q_{e}^{2}}\right) \;\;\;\; (r_{h}\to 0) . 
\label{GMexpand}
\end{eqnarray}
At first glance, it seems that $\phi_{h}$ remains finite in the $r_{h}\to 0$ 
limit in the DEBIon black hole. This is due to the very small diverging rate 
\begin{equation}
\phi_h \sim -\frac{1}{2} \ln(-4\sqrt{bQ_e^2}\ln r_h).  
\end{equation}
This relation is the same as in Eq.~(\ref{asymDEBIonp}). 
The reason for this coincidence is unclear.
The difference in the divergence rate between the GM-GHS and DEBIon 
black holes is a crucial point 
for thermodynamical behaviors and non-existence of the 
extreme solution.  

We are also interested in the scalar field outside or inside the horizon.
By integrating from the event horizon to spatial infinity,
we can find monotonic behavior in the scalar field
for a DEBIon black hole, which has a common property
with the GM-GHS solution, as we showed in Sec. II.
Even small black holes have a structure of the dilaton field
which spreads out to infinity. This is related to the fact that the
dilaton charge $\Sigma$ can be defined by
\begin{eqnarray}
e^{-2\phi}\sim 1+\frac{2\Sigma}{r}, \;\;\; \ \ (r\to \infty).
\label{schargeBI}
\end{eqnarray}
Though the dilaton field has a similar structure near infinity 
in both cases, independently of the size of the
event horizon, it has an intrinsic difference at the small scale.
It behaves as Eq.(\ref{GMexpand}), which is not restricted at the 
horizon but holds everywhere in small scale here, and 
Eq.(\ref{asymDEBIonp}) 
for the GM-GHS solution and the DEBIon black hole, respectively.

\begin{figure}
\begin{center}
\singlefig{10cm}{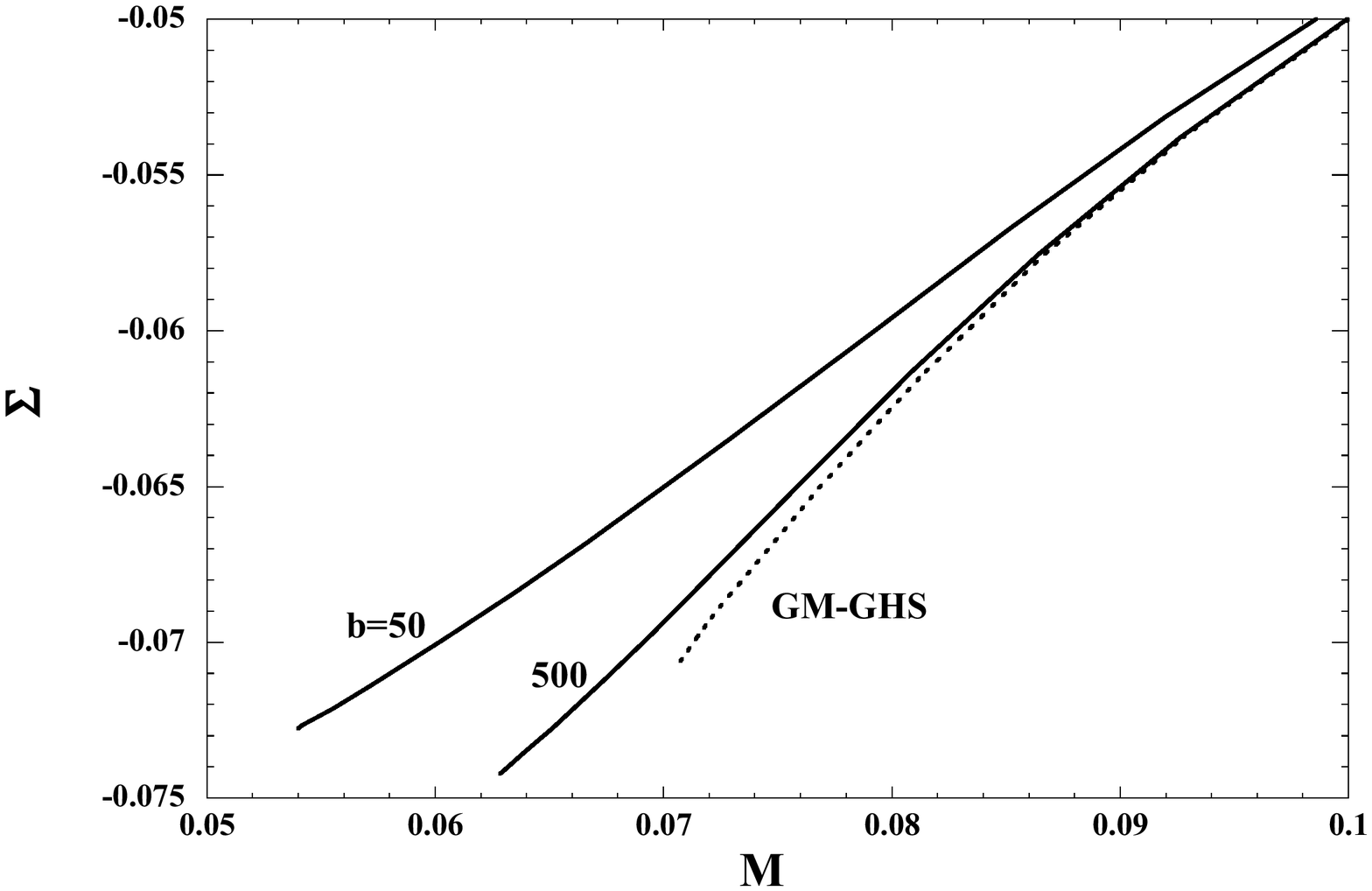}
\caption{$M$-$\Sigma$ diagram of electrically charged 
DEBIon black holes (solid lines) with 
$Q_e =0.1$ and $b=50$ and $500$. 
The GM-GHS solution is plotted by a dotted line. 
This diagram shows that the relation (\ref{dchargeGM}) for 
GM-GHS solution is violated for finite $b$. 
\label{M-schargeele}  }
\end{center}
\end{figure}

Though the small scale behavior of the dilaton field is 
different from the GM-GHS case, we may wonder if 
relation (\ref{dchargeGM}) 
holds even for the DEBIon black hole. 
We show the relation $M$-$\Sigma$ of the electrically charged 
DEBIon black holes (solid lines) with $Q_e =0.1$ and $b=50$ and $500$ 
in Fig. \ref{M-schargeele}. We also plotted the GM-GHS solution by a 
dotted line. This diagram shows that the relation (\ref{dchargeGM}) is 
violated for finite $b$ and absolute value of 
the dilaton charge for the DEBIon black hole 
is larger than that for the GM-GHS solution in the $r_{h}\to 0$ limit. 
Moreover, $|\Sigma |$ becomes larger than the gravitational mass $M$ in 
this limit and the solution does not correspond to 
the BPS saturated state, i.e., 
\begin{eqnarray}
M^{2}+\Sigma^{2}=Q_{e}^{2}. 
\label{BPS?}
\end{eqnarray}
This is the same result in Ref.\cite{Clement} considered 
in the slightly different model from ours. 

\begin{figure}
\begin{center}
\singlefig{10cm}{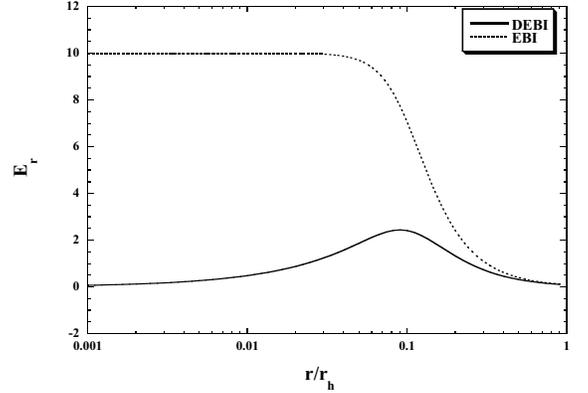}
\caption{The distribution of the electric field $E_r$ diagram of electrically charged 
DEBIon black hole (solid line) and EBIon black hole (dotted line) 
for $r_{h}=0.4$ and $b=100$. Note that the central part of the DEBIon black hole is 
neutral because of the dilaton field. This is one of the reasons why the solution has
a Schwarzschild type global structure. 
\label{r-elein} }
\end{center}
\end{figure}

As we showed in Sec.~II, there is no inner horizon. 
By integrating from the event horizon toward the origin 
with suitable boundary conditions, we can examine the 
internal structure of the black hole. The dilaton 
field monotonically decreases and diverges 
as $\sim \ln r$. The electric field vanishes, approaching 
to the origin by the divergence of the dilaton field, as we show in 
Fig. \ref{r-elein}. We choose $b=100$ and $r_{h}=0.4$. 
The mass function $m$  also diverges as  $\approx r^{-x}$,
$(0<x<1)$. Hence, the function $f$ does not have zero 
except at the event horizon. As a result, the global structure 
is Schwarzschild type. 

\begin{figure}
\begin{center}
\singlefig{10cm}{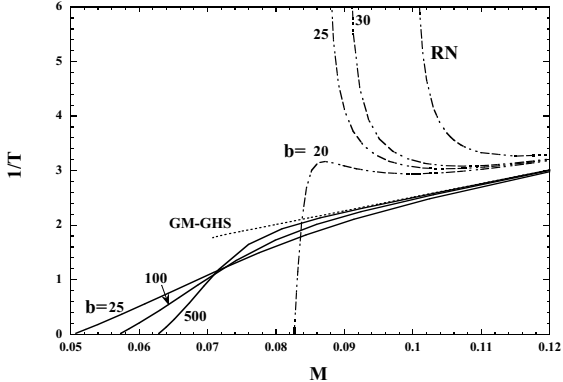}
\caption{$M$-$1/T$ diagram of electrically charged
DEBIon black holes (solid lines) and EBIon black holes (dot-dashed lines)
with the same parameters in Fig.~\ref{M-rhele}.  
\label{M-1Tele} }
\end{center}
\end{figure}

By using Eq.~(\ref{m1}), the temperature of the DEBIon black holes 
is expressed by
\begin{eqnarray}
T& =& \frac{e^{-\delta_h}}{4\pi r_h}(1+2U_h).
\label{temperature}
\end{eqnarray}
It was shown that the thermodynamical properties change drastically in the 
$b \to \infty$ limit by putting the dilaton field on the black hole as seen in 
Fig.~\ref{M-1Tele}. The DEBIon black hole always has 
a higher temperature than the GM-GHS solution 
($T=1/8\pi M$) because of the nonlinearity of the BI field. 
Since the EBIon black hole has the extreme limit,  the temperature becomes
zero in this limit for 
$\alpha_e>\alpha^{\ast}$. On the other hand, 
since there is no extreme solution for DEBIon 
black holes, their  temperature does not vanish but diverges in the 
$r_h \to 0$ limit. 
Hence the evolution by the Hawking
evaporation does not stop until 
the singular solution with $r_h \to 0$ when 
the surrounding matter field does not exist. 

\section{Magnetically charged solution}
EBIon and GM-GHS black holes with magnetic charge can be obtained 
from their electrically charged counterparts owing to 
duality as $F \to \tilde{F}$, and 
$F \to e^{-2\phi}\tilde{F}$, $\phi \to -\phi$, respectively. 
Hence the relations $M$-$r_{h}$ and $M$-$1/T$ never change. 
As is noted above, however, our DEBIon system has no such duality. 
Then it is natural to expect that the magnetically charged solutions have
different properties from electrically charged ones discussed in the
previous section. 

\begin{figure}
\begin{center}
\singlefig{10cm}{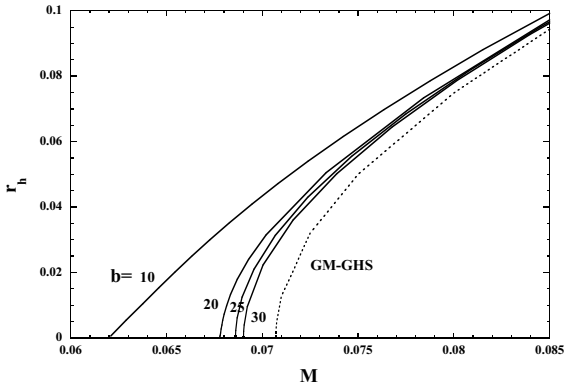}
\caption{$M$-$r_h$ diagram of magnetically charged
DEBIon black holes (solid lines)  with
magnetic charge $Q_m =0.1$ and $b=10$, $20$, $25$ and $30$. 
The GM-GHS black hole is 
also plotted by a dotted line. 
\label{M-rhmag} }
\end{center}
\end{figure}

\begin{figure}
\begin{center}
\singlefig{10cm}{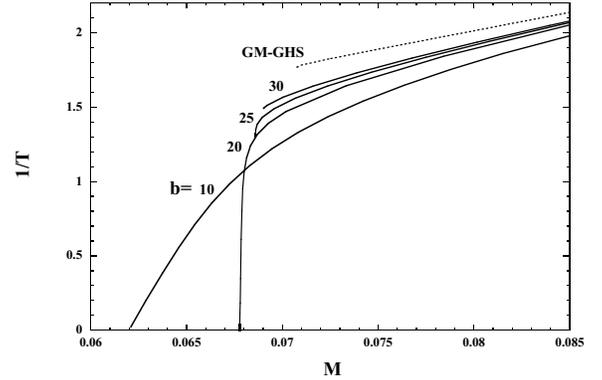}
\caption{$M$-$1/T$ diagram of magnetically charged
DEBIon black holes (solid lines)
with the same parameters in Fig. \ref{M-rhmag}. 
\label{M-1Tmag} }
\end{center}
\end{figure}

We show $M$-$r_h$ and $M$-$1/T$ relations in Fig.~\ref{M-rhmag} 
and Fig.~\ref{M-1Tmag}, respectively. 
Although the $M$-$r_h$ relation seems similar to that of  
the GM-GHS solution regardless of $\alpha_m$, the $M$-$1/T$ relation 
is different depending on $\alpha_m$. 
This is due to the qualitative difference of the behavior 
of the dilaton field in the $r_h \to 0$ limit. For 
$\alpha_m \geq\alpha^{\ast}$, the dilaton field diverges 
at the horizon as 
$\phi_h \sim \ln r_h$ and $\delta_h \sim -\ln r_h$, and 
hence the temperature remains finite as in the GM-GHS case. 
On the other hand, $\phi_h$ and $\delta_h$ are finite 
for $\alpha_m <\alpha^{\ast}$. As a result, 
the temperature diverges. Note that the corresponding solution  
is the particle-like solution. 

\begin{figure}
\begin{center}
\singlefig{10cm}{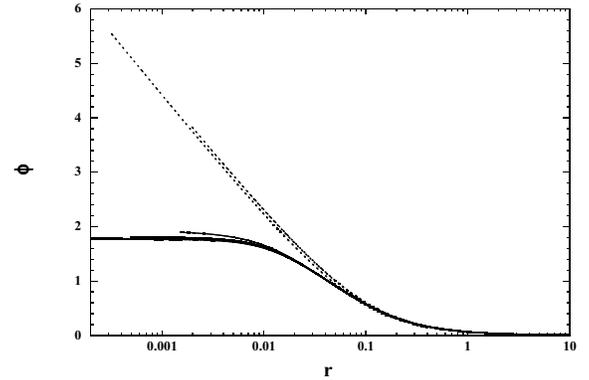}
\caption{The field distribution $r$-$\phi (r)$ of magnetically charged 
DEBIon and DEBIon black holes with $Q_{m}=0.1$ and $b=20$ by solid lines. 
The horizon radii are choosen as $r_{h}=1.685\times 10^{-4}$, 
$4.918\times 10^{-4}$ and $1.499\times 10^{-3}$. 
We also plotted DEBIon black holes with $b=30$ by dotted lines. 
The horizon radii are choosen as $r_{h}=3.522\times 10^{-4}$, 
$1.996\times 10^{-3}$ and $9.564\times 10^{-3}$. 
When the particle-like solution exists, $\phi_{h}' \to 0$ 
in the $r_{h}\to 0$ limit 
and $\phi$ does not diverge. 
\label{r-phimag} }
\end{center}
\end{figure}

We show the field distribution $r$-$\phi (r)$ of DEBIon and DEBIon 
black holes with $Q_{m}=0.1$ and $b=20$ in Fig.~\ref{r-phimag}. 
For reference, we also plot the corresponding DEBIon black hole 
with $b=30$ by dotted lines. Note that while $\phi_{h}'\to 0$ 
and $\phi_{h}$ 
remains finite in the $r_{h}\to 0$ limit with $b=20$, they do not when $b=30$. 
This is the most crucial difference which determines whether 
a particle-like solution exists or not. 
Since the dilaton field is finite
everywhere, the particle-like solution is also relevant in 
the original string frame.
Hence, in the magnetically charged case, 
DEBIon solution {\it does} exist for $\alpha_m <\alpha^{\ast}$. 

\begin{figure}
\begin{center}
\singlefig{10cm}{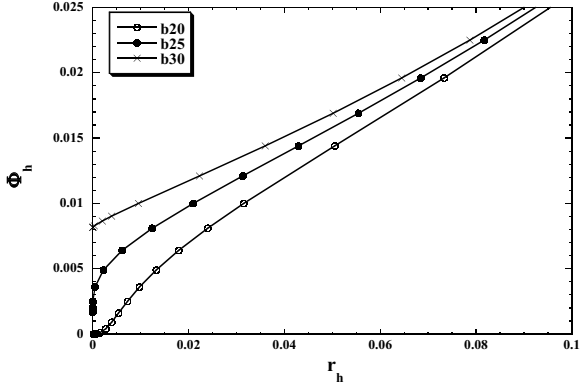}
\caption{$r_{h}$-$\Phi_{h}$ diagram of magnetically charged 
DEBIon black holes for $b=20$, $25$ and $30$. This shows an intrinsic difference 
between $b=25$ and $30$ which determines whether
the extreme solution exists or not . For $b=20$, 
since $\phi_{h}$ itself converges to 
nonzero value in the $r_{h}\to 0$ limit, $\Phi_{h}$ behaves 
as $\propto r_{h}^{2}$ in this limit. 
\label{rh-rh2e2ph-mag} }
\end{center}
\end{figure}

We want to know whether or not the solutions in the $r_{h}\to 0$ 
limit correspond to the extreme ones. From Eq. (\ref{m1}), 
\begin{eqnarray}
\alpha_{m}=\alpha_{m}^{ext}:=\sqrt{\frac{1}{4}+b \Phi_h}
\label{ext}
\end{eqnarray}
for the extreme solutions. Here $\Phi_h := r_{h}^{2}e^{2\phi_{h}}$. 
Thus, there is no extreme solution for $\alpha_m <1/2$. 
For $\alpha_m \geq 1/2$, we must survey $\Phi_h$ (i.e., $\phi_{h}$). 
We show the relation $r_{h}$-$\Phi_h$ 
in Fig.~\ref{rh-rh2e2ph-mag}. 
For $\alpha_m =\alpha^{\ast}$, because 
$\Phi_h \to 0 $ as $r_h \to 0$, the extreme solution is realized 
in the $r_h \to 0$ limit, i.e., $\alpha_{m}=\alpha_{m}^{ext}$. 
For $\alpha_m >\alpha^{\ast}$, $\Phi_h \to const.\neq 0$ as 
$r_h \to 0$. We cannot tell whether Eq.~(\ref{ext}) is fulfilled 
for a certain horizon radius since $\phi_h$ is obtained
iteratively only by numerical method. Our calculation always shows 
$\alpha_{m}<\alpha_{m}^{ext}$ except for $\alpha_{m}=\infty$, 
which implies that the extreme solution is realized only when 
$\alpha_{m}=\alpha^{\ast}$ and $\alpha_{m}=\infty$. 
We also show the result for $\alpha_m <\alpha^{\ast}$ for reference. 
In this case, $\Phi_{h}\propto r_{h}^{2}$ in the $r_{h}\to 0$ limit. 
This shows that $\phi_{h}$ converges to finite value in this limit. 

As we considered in the electrically 
charged case, we can divide the contributions from the dilaton field to 
the mass function. They are, however, more complicated than in 
the electrically charged 
case: (i) and (ii) are the same and there is another factor
(iii) the prefactor $e^{-4\phi}$ before $Q_{m}^{2}$ in Eq.~(\ref{BI}). 
Since  both factors (i) and (ii) 
make positive contributions to the mass function because $\phi >0$, 
one may think that the gravitational mass 
becomes larger than the corresponding EBIon and EBIon black hole. But 
this is not the case. The third factor makes that the effect of 
the magnetic charge reduce 
by $e^{-4\phi}$, which forces the solution to approach the Schwarzschild one. 
As a result, the gravitational mass is reduced compared to the 
EBIon black hole. 

\begin{figure}
\begin{center}
\singlefig{10cm}{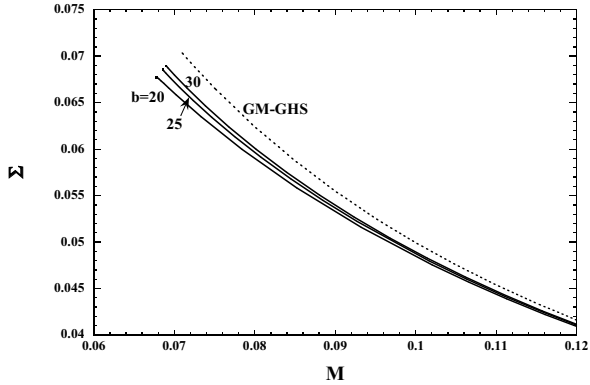}
\caption{$M$-$\Sigma$ diagram of magnetically charged 
DEBIon black holes with $Q_{m}=0.1$ and $b=20$, $25$ and $30$ by solid lines. 
We also plotted the GM-GHS solution by dotted line. 
\label{M-schargemag} }
\end{center}
\end{figure}

We may think the effect of the dilaton field also makes a 
qualitative difference for the dilaton charge. We show the $M$-$\Sigma$ 
relation of a DEBIon black hole with $Q_{m}=0.1$ and $b=20$ and $30$. 
in Fig.~\ref{M-schargemag}. 
We can find that a qualitative difference 
near the horizon does not cause strong influence on the asymptotic region. 
Contrary to the electrically charged case, the dilaton charge of the 
DEBIon black hole is smaller than that of the GM-GHS solution in the 
$r_{h}\to 0$ limit. There is no BPS saturated solution 
in this case, either. 

The behaviors of the field functions inside the event
horizon are similar to those in the electrically charged
case qualitatively, except for the sign of the dilaton
field. The magnetic field diverges as $B_r \approx
r^{-x}$, $(1<x<2)$. As we showed in Sec. II, there is no inner horizon 
and the dilaton field monotonically increases toward the origin. 
The nonexistence of the inner horizon is a characteristic feature of 
the black hole solution with a scalar field except for some special cases. 
The equations of the scalar fields become singular on the horizon where 
$f=0$, and the scalar field must satisfy a certain boundary condition there. 
However, when we integrate the scalar field equation from the black hole 
event horizon inward, this boundary condition can not be fulfilled 
in general and the scalar field diverges. As a result, there is 
no inner horizon. 
It may seem strange that there should be an extreme solution in spite 
of the fact that there is no inner horizon. In the EMD system, this can be 
understood as follows. The GM-GHS solution with the monopole 
charge has no inner horizon. But if we consider the dyon case 
in the EMD system, i.e., trivial axion, there appears an inner horizon. 
This corresponds to the special case. 
The inner horizon shrinks to $r\to 0$ in the limiting case, 
i.e., $Q_{e}=0$ or $Q_{m}=0$. 
The inner horizon and the event horizon can degenerate at $r=0$ when 
$M=\sqrt{2}Q_{e(m)}$. 
Thus, the extreme solution for the electrically 
(or magnetically) charged solution appears. 
In our case, we may think that 
only the magnetically charged solution corresponds to such a case. 
 

\section{dyon solution}

Finally, we investigate the dyon solutions which have both electric and 
magnetic charges. The dyon solution 
is important for the following reasons.
In the EMD system, the black hole solutions have different properties from
the monopole case. For example, there is an inner horizon and the 
global structure is the RN type\cite{GM-GHS,Cheng}. The inner horizon 
shrinks to zero in the monopole ($Q_{m}\to 0$ or $Q_{e}\to 0$) limit. 
The temperature becomes zero in the extreme limit 
while it is finite in the monopole cases. 
On the contrary, the dyoic black hole solution in the EMDA system 
can be obtained from the monopole one because of the SL($2$,R) duality 
\cite{Wilczek}: 
\begin{eqnarray}
\lambda \to \frac{a\lambda +b}{c\lambda +d},\ \ \ \ 
ad-bc=1, \nonumber  \\  
F_{+}\to -\lambda F_{+},\ \ \ \ 
F_{-}\to -\bar{\lambda} F_{-},
\label{SL(2,R)F}
\end{eqnarray}
where $\lambda :=a+ie^{-2\phi}$, $F_{\pm}:=F\pm i\tilde{F}$ and 
$\bar{z}$ means the complex conjugate of $z$. This is why the 
temperature and inner structure coincide with the monopole case. 
We expect similar properties in the modified EBIDA model 
which has SL($2$,R) duality. 
However, since we treat the model without such duality, we can expect 
that nontrivial changes occur in our system 
and that the monopole solutions are 
its special cases. Moreover, we are interested in the effect of 
the axion field which was not investigated before. 
Since we find a trivial value for the monopole solutions under 
the static and spherically symmetric ansatz, we should by all means 
investigate the dyon solution. It may result in significant differences 
in such aspects as the behavior of the dilaton field 
or in the temperature. 

\begin{figure}
\begin{center}
\segmentfig{10cm}{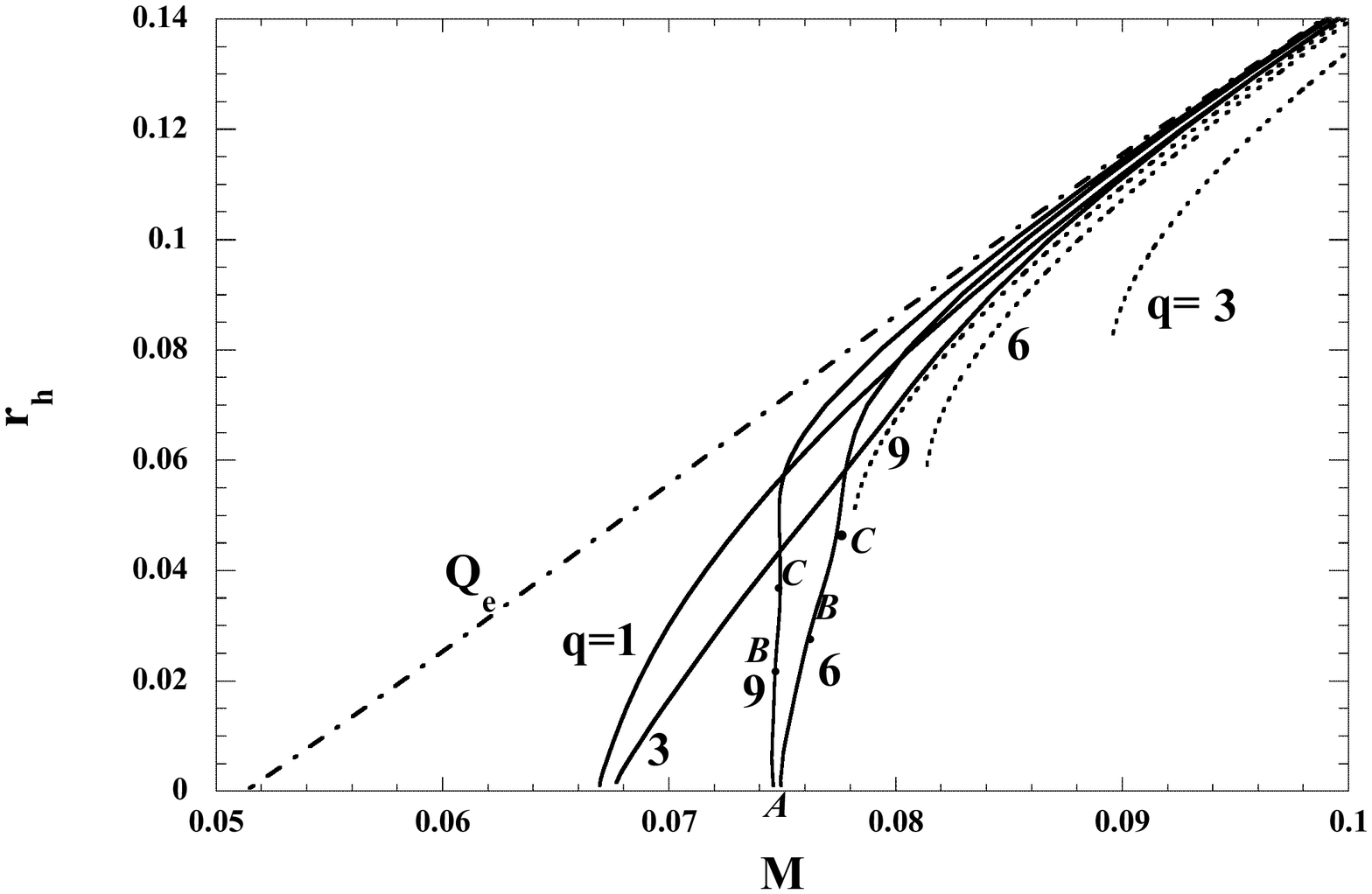}{(a)}  \\
\segmentfig{10cm}{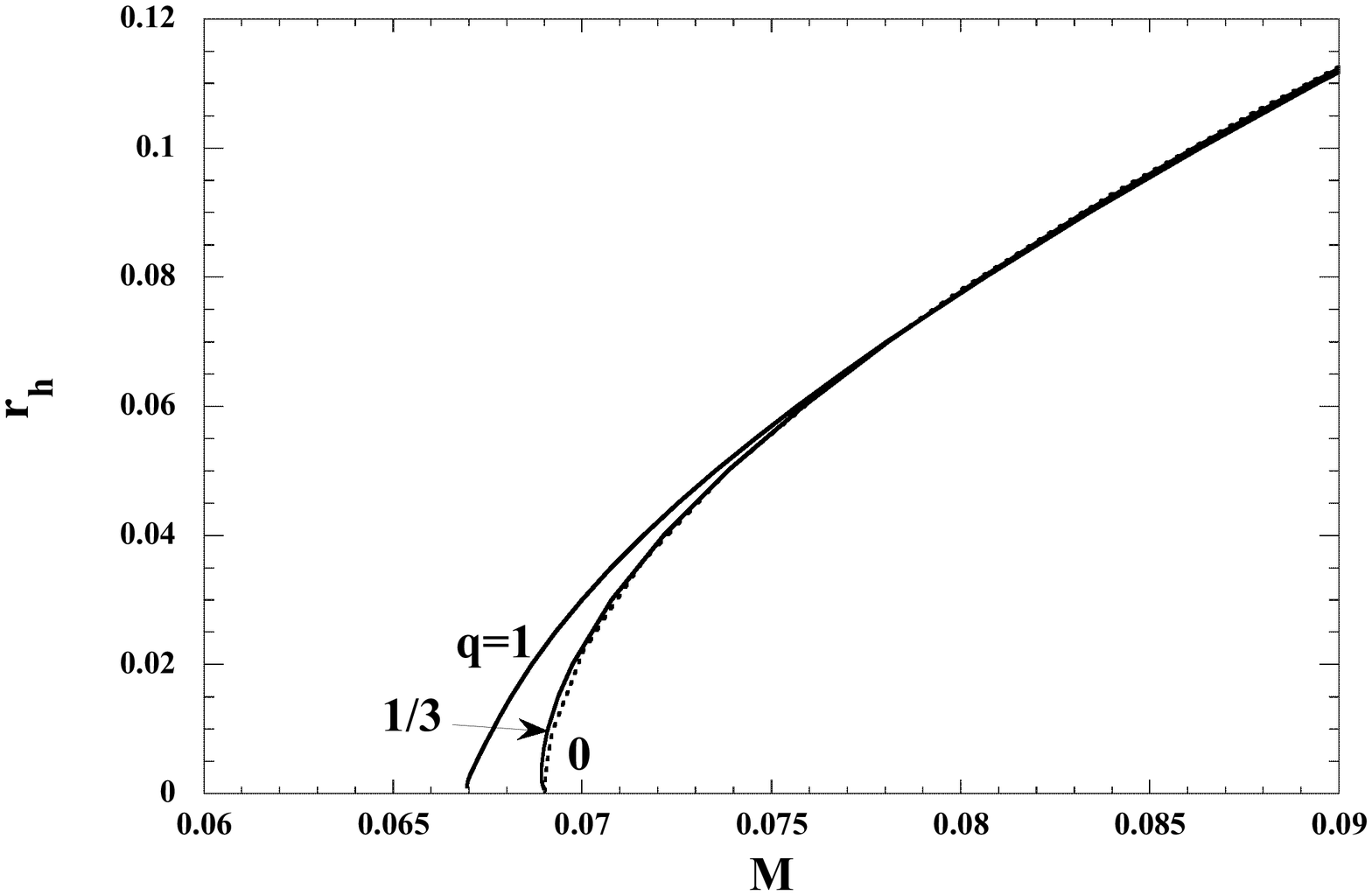}{(b)}
\caption{$M$-$r_{h}$ diagram of the dyon solutions 
with $\sqrt{Q_{m}^{2}+Q_{e}^{2}}=0.1$ and $b=30$. 
We plot the solutions $q\geq 1$ in (a) and $q\leq 1$ in (b). 
We can find that though the solutions with small $q$ approach the magnetically 
charged case (i.e., $q\to 0$), those with large $q$ do not approach the electrically 
charged case (i.e., $q\to \infty$) because of nontrivial distribution 
of the axion field. We also plot the dyon solutions in the EMD system by 
dotted lines in (a). \label{M-rhdyon} }
\end{center}
\end{figure}

\begin{figure}[htbp]
\singlefig{10cm}{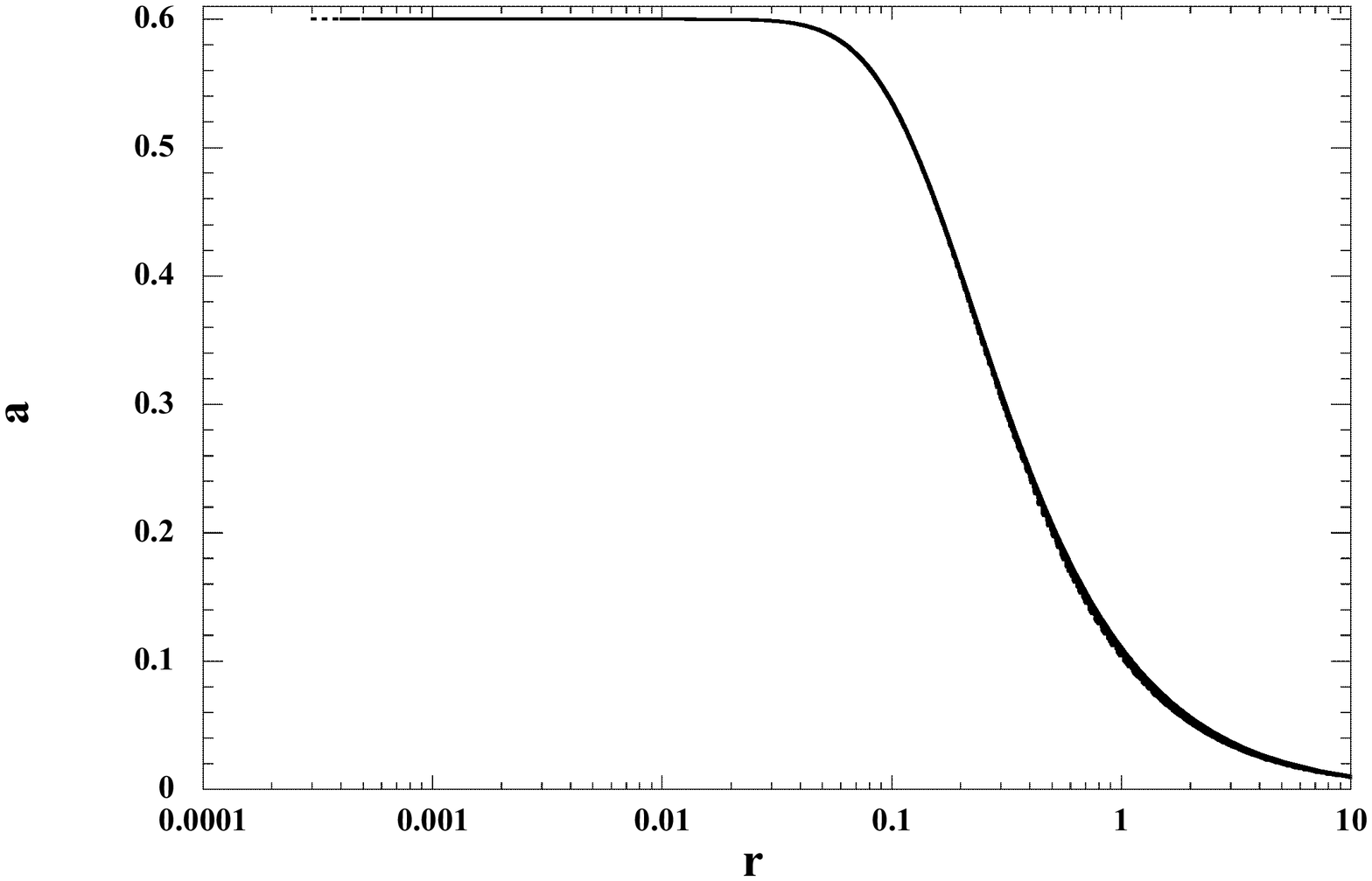}{}
\caption{The field distributions $r$-$a (r)$ 
for $b=30$ and $q=0.6$ in numbers of dyon solutions for 
$3\times 10^{-4}\leq r_{h}\leq 5\times 10^{-3}$. 
They are almost indistinguishable, which means that 
the distribution of the axion field is almost independent of the horizon radius. 
\label{r-?dyon}}
\end{figure}

We first show the $M$-$r_{h}$ relation of dyon solutions with 
$\sqrt{Q_{e}^{2}+Q_{m}^{2}}=0.1$ and $b=30$. We plot the solutions 
with $q\geq 1$ 
in Fig.~\ref{M-rhdyon} (a). Since dyon solutions in the EBI system 
and in the EMDA system have the same 
$M$-$r_{h}$ relations, we omit them. We also plot the 
solutions $q\leq1$ in Fig.~\ref{M-rhdyon} (b). 
In the EMD system, solutions approach the 
electrically charged (or magnetically charged) ones 
in the $q\to \infty$ (or $\to 0$) limit. 
We can find, however, that the solutions with large $q$ {\it do not} approach 
the electrically charged case while those with small $q$ approach the magnetically 
charged case. This is due to the 
nontrivial distribution of the axion field. 
To see this, we investigate the scalar field of the small black hole solutions. 
Our numerical analysis shows that while $\phi_{h}$ increases as $r_{h}$ decreases for 
$r_{h}\gsim 10^{-3}$, it 
decreases in the $r_{h}\to 0$ limit and $\phi_{h}'$ approaches zero. 
This may seem to be the symptom of the existence 
of the particle-like solution as in the magnetically charged case. 
However, this is not the case. We show the configurations of the axion field 
for $b=30$ and $q=0.6$ in numbers of solutions 
for $3\times 10^{-4}\leq r_{h}\leq 5\times 10^{-3}$ in Fig. \ref{r-?dyon}. 
We can see that solutions have the same scale structure 
almost independently on the horizon radius, 
and there will be a solution with a non-trivial axion field in the $r_{h}\to 0$ 
limit. However, it does not correspond to the particle-like solution. 
The regularity at the origin requires $a(0)=q$ and $a'(0)=0$ by Eq. (\ref{a1}), 
which means that $a$ is constant $a\equiv q$. We can confirm that there is such 
a solution with $a\equiv q$ and the non-trivial dilaton field. It incorporates 
with the boundary condition $a(\infty )=0$. However, the asymptotic flatness 
requires just $a(\infty )=$const. We choose $a(\infty )=0$, since if the 
asymptotic value of $a$ is zero at the initial condition before the gravitational 
collapse, it can not be changed by a physical process with finite energy. 
Hence the condition $a(\infty )=0$ is assumption. If $a(\infty)\neq 0$ at initial, 
we should choose a different boundary condition. In this sense, the above solution 
with $a\equiv 0$ is also the particle-like solution in system (\ref{EBID}) 
with a suitable boundary condition. It should be noted, however, that this 
solution is formed only when the initial data satisfies $a(\infty )=q$. 
Hence, this solution is not generic if we consider the formation process 
of the solution. 

\begin{figure}
\begin{center}
\singlefig{10cm}{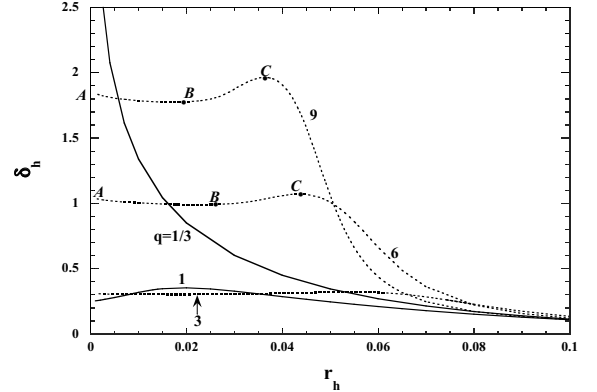}
\caption{$r_{h}$-$\delta_{h}$ relation for the dyon solutions with $b=30$. 
Though $\delta_{h}$ diverges in the $r_{h}\to 0$ limit for $q=1/3$, 
it does not for other cases. For other values of $q$, we can find that $\delta_{h}$ 
decreases as $r_h$ decreases, but it eventually increases in the $r_{h}\to 0$ 
limit because of the large contribution from the axion field. 
\label{rh-delhdyon} }
\end{center}
\end{figure}

To understand the rather complicated $M$-$r_{h}$ diagram, 
we focus on the metric function $\delta$ 
which represents the contribution from the gradient terms of 
the axion and the dilaton fields. 
We show the $r_{h}$-$\delta_{h}$ relation 
for $b=30$ in Fig. \ref{rh-delhdyon}. 
$\delta_{h}$ diverges in the $r_{h}\to 0$ limit for $q=1/3$ 
since $\phi_{h}$ diverges. For large values of $q$, (e.g., $q=6$, $9$), 
the behavior is complicated. As $r_{h}$ decreases from the point $C$ 
to $B$, $\delta_{h}$ also decreases because the contribution from 
the dilaton field becomes small. However, as $r_{h}$ decreases further 
from $B$ to $A$, $\delta_{h}$ increases because of the large contribution 
from the axion field. 
This behavior is crucial for the solutions not to approach the electrically 
charged solution in the $q\to \infty$ limit. 
These behaviors of $\delta$ are reflected in the $M$-$r_{h}$ diagram. 
The points $A$, 
$B$ and $C$ in the $M$-$r_{h}$ diagram correspond to these in the 
$r_{h}$-$\delta_{h}$ diagram. The curves from $C$ to $B$ 
become very steep because 
of the decreasing contribution from the gradient term of the scalar fields 
to the gravitational mass $M$. This tendency is relaxed from $B$ to $A$. 
Thus, the rather complicated curves in the $M$-$r_{h}$ diagram 
result from the contributions of two scalar fields.  

\begin{figure}
\begin{center}
\segmentfig{10cm}{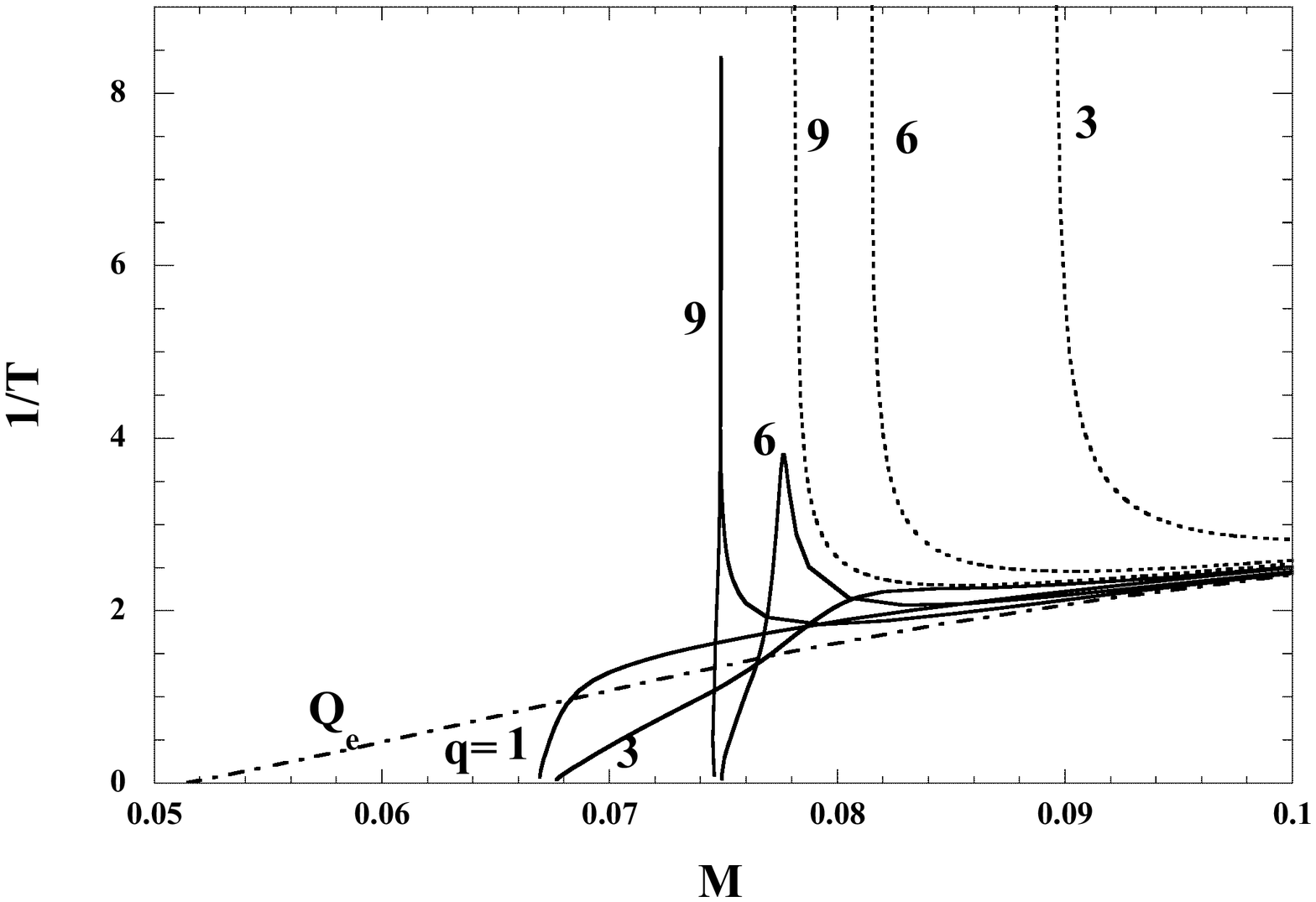}{(a)}  \\
\segmentfig{10cm}{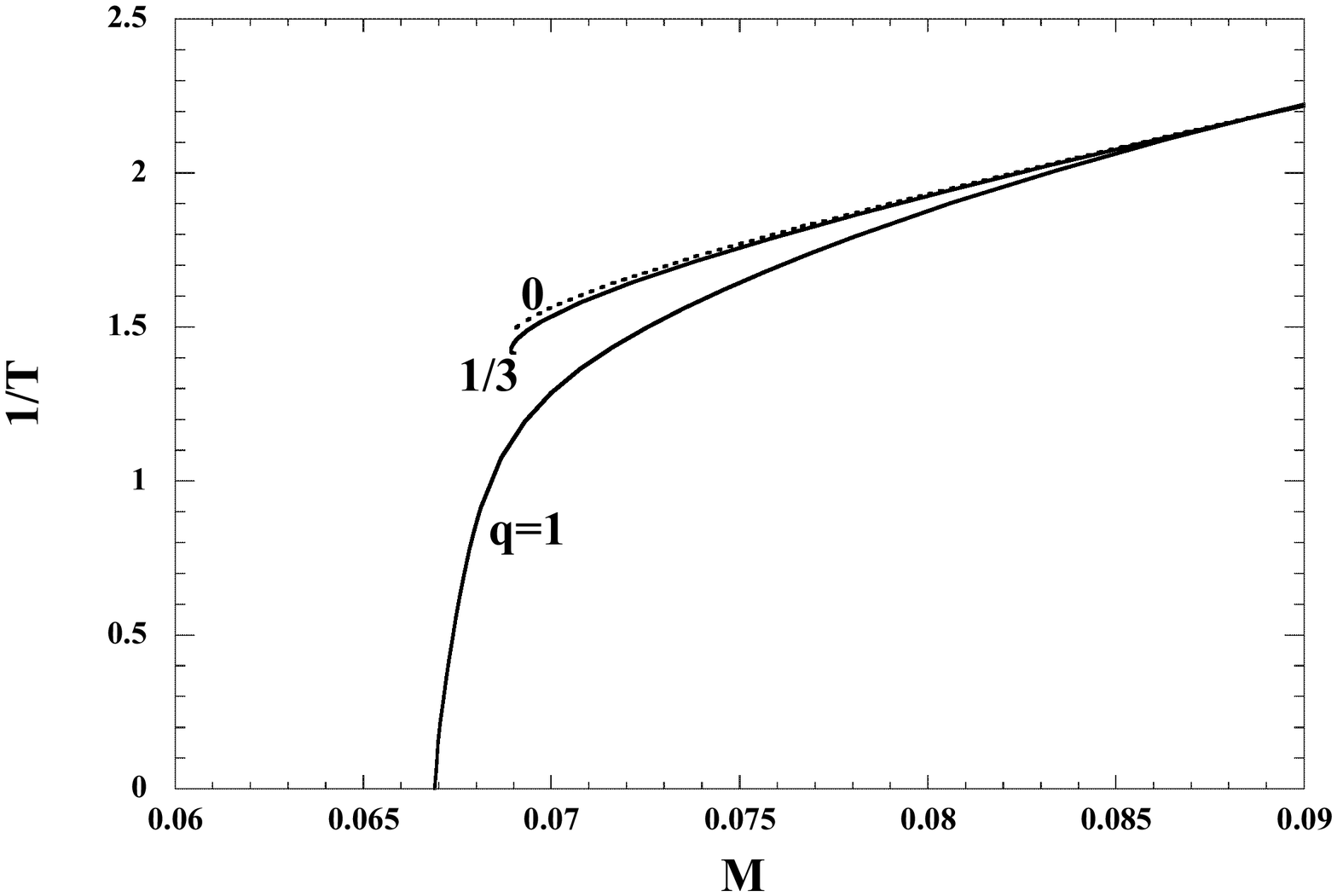}{(b)}
\caption{$M$-$1/T$ diagram with the same parameters in Fig. \ref{M-rhdyon}. 
We can find that the temperature diverges for $q\geq 1$ which can be 
interpreted in the asymptotic analysis. For large $q$, though the temperature 
approaches zero once, it does not have zero limit in the $r_{h}\to 0$ but 
diverges in this limit. On the contrary, the dyon solutions in the EMD system 
have an extreme limit where the temperature becomes zero. 
\label{M-1Tdyon} }
\end{center}
\end{figure}

We show the $M$-$1/T$ relation in 
Fig. \ref{M-1Tdyon}. GM-GHS solutions 
have zero temperature limit. While the solutions in the EBIDA system 
for large $q$ show similar behavior to the GM-GHS solutions 
in the large mass region, their temperature does not vanish 
in the $r_{h}\to 0$ limit, but instead diverges. 
For small $q$ ($q=0$, $1/3$), the temperature remains finite. 
These properties are understood by the following considerations. 

As we showed above, $a_{h}'$ must be negative and $a$ must be a monotonically 
decreasing function to satisfy $a(\infty)\to 0$, since $a$ cannot have a 
local minimum. We can find from Eq. (\ref{ath}) that $a_{h}'\to -\infty$ 
in the $r_{h}\to 0$ limit if both $e^{-2\phi_{h}}$ and $Y_{h}$ do not 
approach zero. Our numerical calculation shows that $Y_{h}$ approaches zero 
fast enough to satisfy $Y_{h}/r_{h}^{2}\to 0$ in the $r_{h}\to 0$ limit. 
Because of this property, we obtain $a_{h}'\to 0$ in the $r_{h}\to 0$ limit. 
Thus, we can write it as 
\begin{eqnarray}
\lim_{r_{h}\to 0}m'(r_{h})=\alpha_{m}  .
\label{thermo-condition}
\end{eqnarray}
Whether the temperature diverges or not in the $r_{h}\to 0$ limit 
is determined only by $\alpha_{m}$ as we see from Eq. (\ref{temperature}). 
As in the magnetically charged case, for 
$\alpha_m \geq\alpha^{\ast}$, the dilaton field on the horizon diverges as 
$\phi_h \sim \ln r_h$ and $\delta_h \sim -\ln r_h$, and 
hence the temperature remains finite as in the GM-GHS case. 
On the other hand, $\phi_h$ and $\delta_h$ are finite in the 
$r_h \to 0$ limit for $\alpha_m <\alpha^{\ast}$. As a result, 
the temperature diverges. 
If we restrict it as $\sqrt{Q_{e}^{2}+Q_{m}^{2}}=0.1$, this condition is 
expressed as 
\begin{eqnarray}
\frac{b}{1+q^{2}}<25  .
\label{thermo-condition2}
\end{eqnarray}
For $b=30$, this inequality becomes $q>1/\sqrt{5}\approx 0.4\ldots$ 
and is consistent with our result 
shown in Fig. \ref{M-1Tdyon}. 
Though we showed only one example, we confirmed this behavior 
for various parameters. 
It is interesting to compare this result with 
the corresponding property of 
the dyon solution in the EMDA system where the dilaton field in the 
$r_{h}\to 0$ limit approaches the magnetically charged case while 
the relation $T=1/(8\pi M)$ always holds because of the 
SL($2$,R) duality. 

Let us consider the evaporation process for $q\gg1$. 
There is a point where $d(1/T)/dM=0$ is satisfied and has 
very low temperature. When the black hole at this point loses its mass 
a little, we can see a rapid growth in the temperature 
which may cause an explosion. What brings about this property? 
Considering the analogy of 
the EMD and the EMDA system, it is plausible that 
for the dyon solution, extreme solution in the EBID system may appear 
at the mass scale 
where the temperature becomes zero. So the temporal approach to 
zero temperature will be interpreted as an effect of the dyon. 
On the other hand, if we include the axion, we can prove that there is 
no inner horizon. So we can find that there is a solution in the 
$r_{h}\to 0$ limit as in the EMDA system. Taking these things into account, 
the property shown above is caused by the combination of the dyon 
with the axion field. We may see that if the 
EBIDA system is chosen to satisfy SL($2$,R) invariance, 
there is no such property. 


\section{Discussion}
We investigate the static spherically symmetric solutions in the 
EBIDA system. When the solutions 
do not have both electric and magnetic charge, there is no contribution 
from the axion field. (i) In the electrically charged case, there
is neither extreme solution nor particle-like solution, both of which
exist in the EBI system. The temperature of the black hole 
diverges monotonically 
in the $r_h \to 0$ limit. The dilaton charge $\Sigma$ becomes larger 
than the gravitational mass $M$ in this limit. (ii) In the 
magnetically charged case, the extreme 
solution exists when $\alpha_m =1/2$ and particle-like solutions exist when 
$\alpha_m <1/2$. For $\alpha_m >1/2$, the solution 
in the $r_h \to 0$ limit 
corresponds to naked singularity. The temperature of the black hole 
remains finite in the $r_h \to 0$ limit for $\alpha_m \geq 1/2$, 
while it diverges for $\alpha_m <1/2$. There is no BPS saturated solution. 
(iii) In the dyon case, we can obtain a nontrivial axion field. 
Contributions of the axion field mainly depend on the ratio 
$q=Q_{e}/Q_{m}$. Since $a_{h}$ is restricted to 
$a_{h}<q$, solutions in the $q\to 0$ limit approach the corresponding 
magnetically charged solutions. But they do not approach the 
electrically charged ones in the $q\to \infty$ limit. 
As for the temperature, whether or not it diverges in the 
$r_{h} \to 0$ is only determined by $\alpha_{m}$. 
We can show that there is no inner horizon in any charged case, 
and the global structure is the Schwarzschild type. 

Here, we discuss some outstanding issues. 
First, we compare our results with those in Ref. \cite{Clement} 
where a slightly different system from ours was considered. 
What they call the soliton 
corresponds to our ``particle-like" solution considered in the 
electrically charged case. We required the finiteness of the 
dilaton field for a particle-like solution and 
find such a solution in the magnetically charged case. 
They did not investigate the dyon case. However, since the extended version 
of the EBIDA model has a SL(2,R) duality, it is easy to produce the 
dyon solutions. Those solutions will have distinct 
properties from us. 

Second, we comment on the stability of solutions and the relation between 
the inverse string tension $\alpha '$ and the gravitational constant $G$. 
We considered them in our previous paper on the monopole case\cite{TTpast}. 
By using catastrophe theory\cite{cata,torii}, 
we find that the discussion is irrelevant even if we include the dyon case. 
This implies that our solutions are stable against spherical perturbations. 
As we discussed following Ref. \cite{Gibbons}, we identify the supersymmetric 
spin $0$, $1/2$ particle with the extreme solution. 
For the dyon case, the extreme solution is realized for 
$b=25(1+q^{2})$ when we fixed the charge $\sqrt{Q_{e}^{2}+Q_{m}^{2}}=0.1$. 
We can find in Fig. \ref{M-rhdyon} that the result is not affected from the 
monopole case. We find $2\pi\alpha\sim 1.73G$ for the 
dyonic DEBIon black hole. 

Finally, we denote future work. 
Though we find the particle-like solution in the magnetically charged case, 
it is unsatisfactory as a candidate of the remnant of 
the Hawking radiation for the following reasons. 
(i) Since it appears in the $r_{h}\to 0$ limit where the Hawking temperature 
diverges, the quantum effect of the gravity may affect the results. 
(ii) Since the particle-like solution admits a conical singularity at the 
origin, there still appears naked singularity. 
These results may be modified if 
the higher curvature terms are taken into account. 
As was already 
pointed out, it is difficult to obtain the counterpart of the BI action 
in the gravity part. It is open to question whether or not 
the singularity inside the horizon is regulated as the electric 
field is in the BI action. Concerning this, BI type action for 
gravity was considered in Ref. \cite{Feigen} and the regular black hole solution 
was obtained, though there is no theoretical background at present. 
But considering higher curvature such as the Gauss-Bonnet 
term is still important, since if the black hole solution is not singular 
in this case, there may be a possibility that this result is preserved 
even if we consider the higher curvature term. This type of 
consideration may shed light on the realization of the dream in 
Ref. \cite{dream}. 

\section*{Acknowledgments}
Special thanks to Gary W. Gibbons, Daisuke Ida, Kei-ichi Maeda and 
Shigeaki Yahikozawa for useful 
discussions.  T. T. is thankful for  financial support from the 
JSPS. This work was supported  by a JSPS Grant-in-Aid (No. 106613), and 
by the Waseda University Grant  for Special 
Research Projects.


\begin{thebibliography}{99}
\bibitem{foot2}
By this term we mean a solution which has conical singularity 
at the origin and has finite energy.
\bibitem{notes}
In this paper, we study the tree-level action. When we
are interested in the higher-loop action, the coupling constant $\gamma$
takes different value by assuming anomaly cancellation 
(See. Ref.~\cite{Callan}).
\end{thebibliography}
\end{document}